\documentclass[Journal,letterpaper]{ascelike-new}
\usepackage[utf8]{inputenc}
\usepackage[T1]{fontenc}
\usepackage{lmodern}
\usepackage{graphicx}
\usepackage[figurename=Fig.,labelfont=bf,labelsep=period]{caption}
\usepackage{subcaption}
\usepackage{amsmath}
\usepackage{multirow}
\usepackage{newtxtext,newtxmath}
\usepackage[colorlinks=true,citecolor=red,linkcolor=black]{hyperref}
\usepackage{algorithm}
\usepackage[noend]{algpseudocode}
\usepackage{pdflscape}
\usepackage{makecell}
\NameTag{Agapaki, \today}
\begin{document}
\nolinenumbers

% You will need to make the title all-caps
\title{CLOI: AN AUTOMATED BENCHMARK FRAMEWORK FOR GENERATING GEOMETRIC DIGITAL TWINS OF INDUSTRIAL FACILITIES}

\author[1]{Eva Agapaki}
\author[2]{Ioannis Brilakis}

\affil[1]{Senior Software Developer, PTC Inc.,U.S.A. Email: agapakieva@gmail.com}
\affil[2]{Laing O’Rourke Reader, Department of Engineering, University of Cambridge, CB2 1PZ, U.K.}

\maketitle

% Please include an abstract:
\begin{abstract}
This paper devises, implements and benchmarks a novel framework, named \textit{CLOI}, that can accurately generate individual labelled point clusters of the most important shapes of existing industrial facilities with minimal manual effort in a generic point-level format. \textit{CLOI} employs a combination of deep learning and geometric methods to segment the points into classes and individual instances. The current geometric digital twin generation from point cloud data in commercial software is a tedious, manual process. Experiments with our \textit{CLOI} framework reveal that the method can reliably segment complex and incomplete point clouds of industrial facilities, yielding 82\% class segmentation accuracy. Compared to the current state-of-practice, the proposed framework can realize estimated time-savings of 30\% on average. \textit{CLOI} is the first framework of its kind to have achieved geometric digital twinning for the most important objects of industrial factories. It provides the foundation for further research on the generation of semantically enriched digital twins of the built environment. 
\end{abstract}

\section{Introduction}
The industrial sector and especially the oil and gas is an industry with the highest potential growth in terms of worker productivity and economic value of the sector within the next couple of years. The Global Infrastructure Initiative forecasts that heavy industrial buildings and the oil and gas sector are among the construction sectors with the highest potential for investments with an average Compound Annual Growth Rate (CAGR) of 3.4\% \cite{McKinseyGlobalInstitute2015DigitalHave-Mores}. Therefore, it is crucial that the industrial sector is properly maintained given the high value of the industrial assets for our economies. 

Maintenance, safety management and retrofitting are vital operations in the life-cycle of existing industrial facilities. Corrective or poor maintenance incurs unplanned downtime costs, which are estimated to be \$50 billion per year \cite{NationalInstituteofStandardsandTechnology2018TheManufacturing}. The primary reasons for these incidents are ineffective and inefficient facility management and poor mapping of the existing industrial equipment. Faster digital industrial documentation is urgently required to reduce unscheduled equipment downtimes and boost the Overall Equipment Effectiveness (OEE) of a factory, which is currently estimated to be between 5 to 20\% \cite{PECI1999PortableOperations}.

There are limits on the acceptable shut down duration that will not impede production. These limits cannot be violated without incurring extra costs. This is why adoption of Digital Twins (DTs) is crucial for the industrial sector. The greatest value of using DTs is that they are projected to save substantial costs for facility managers by automating the preventive maintenance process which will enable accurate positioning of each industrial object and timely maintenance decisions. For example, DTs can help to keep records of the inventory, processes, historical data and additional equipment. This allows owners to identify inefficiencies and ways to address them. Studies show that the wider adoption of DTs will unlock 15-25\% savings to the global infrastructure market by 2025 \cite{Barbosa2017ReinventingProductivity,Gerbert2016DigitalConstruction}.

The concept of DTs is not new. NASA first generated the term ``twin'' when building two identical space vehicles for its Apollo program \cite{Glaessgen2012TheVehicles}. The modern terminology of a ``digital twin'' has been attributed to Dr Michael Grieves as part of his research in Product Lifecycle Management (PLM) \cite{Grieves2014DigitalReplication}. Reports based on the digitization index have shown that the oil and gas industry has been highly digitised as compared to the construction industry, which is in the bottom of the list \cite{Agarwal2016TheConstruction}. Despite the high value DTs have in the industrial sector, yet, industrial facilities do not have DTs for existing industrial factories due to the high perceived cost which outweighs their benefits \cite{West2017IsProject}. 

The generation of a geometric Digital Twin (gDT) is the core and first step in the DT generation \cite{borrmann2018principles}. The inputs for the generation of gDTs are usually point clouds scanned with Terrestrial Laser Scanners (TLS) \cite{Marshall2016HandbookScanning}. 90\% of the gDT generation cost is spent on converting point cloud data to 3D models due to the sheer number of objects of each industrial facility \cite{Fumarola2011GeneratingApproaches,Hullo2015Multi-SensorArchitectures}. Hence, cost reduction is only possible by automating the generation of gDTs. However, automatically classifying millions of objects is a very hard classification problem due to the very large number of classes and the strong similarities between them. We provided in our previous work \cite{Agapaki2018PrioritizingFacilities} a comprehensive technical assessment and viable evaluation of existing state-of-the-art software tools available. In the following paragraphs, we summarize the state-of-practice based on this evaluation.

\subsection{State-of-practice}

In our previous work \cite{Agapaki2018PrioritizingFacilities}, we identified the most frequent and laborious to model object types, which are cylindrical objects (straight pipes, electrical conduit and circular hollow sections), valves, elbows, I-beams, angles, channels and flanges. Cylinders require 80\% of the total modelling time of the ten most important object types in EdgeWise \cite{ClearEdge2019PlantCapabilities} and represent 45.5\% of the total number of objects in an industrial plant on average. EdgeWise was selected compared to other state-of-the-art software, because it is the only commercially available tool that attempts to automatically extract cylinders from the point cloud of an industrial plant without significant user assistance. EdgeWise has significantly accelerated 3D modelling of industrial plants according to the findings discussed above. However, it has some limitations, which can be summarized as follows: 

\begin{enumerate}\setcounter{enumi}{0}
  \item Structural elements (I-beams, angles, channels) should be manually modelled and their location in the point cloud is roughly defined based on the modeler's discretion. 
  \item Segmentation of cylinders has been partially achieved with detection rates being 75\% recall and 62\% precision on average \cite{Agapaki2018PrioritizingFacilities}. The same metrics for cylindrical objects labelled as pipes are 58\% and 47\% respectively. It is also important to note that EdgeWise erroneously includes points that do not belong to a geometric shape. This is due to fitting errors, which occur since primitive shapes are perfect shapes, whereas the scanned, physical objects are imperfect (e.g. a cylindrical pipe may be bent). 
  \item EdgeWise is not designed to output geometric shapes in an open and generic format. As such, modelers cannot easily exchange data between different operational-phase gDT platforms due to data inconsistency between them. 
\end{enumerate}

Therefore, the evaluation of EdgeWise uncovered (a) the substantial performance of this software in detecting cylinders with its pitfalls, (b) the inability of the software to (i) further classify cylinders into conduit or pipes or CHSs and (ii) detect and further classify I-beams, channels, flanges, valves and angles in spite of their high frequency in an industrial facility.

This performance of EdgeWise has substantial room for improvement and this paper intends to address the above-mentioned limitations in order to automatically generate gDTs of industrial facilities and assist the tedious current practice. We propose a geometric twinning framework for existing industrial facilities and bench-mark it with the current state of practice. In the following section, the state-of-the-art research methods related to the above-mentioned limitations are presented. We then outline the framework in the proposed solution, which is followed by the experiments and results. The conclusions are then derived in the last section.

\section{BACKGROUND}
\label{background}

\begin{figure}[!ht]
\centering
\includegraphics[width=0.9\textwidth]{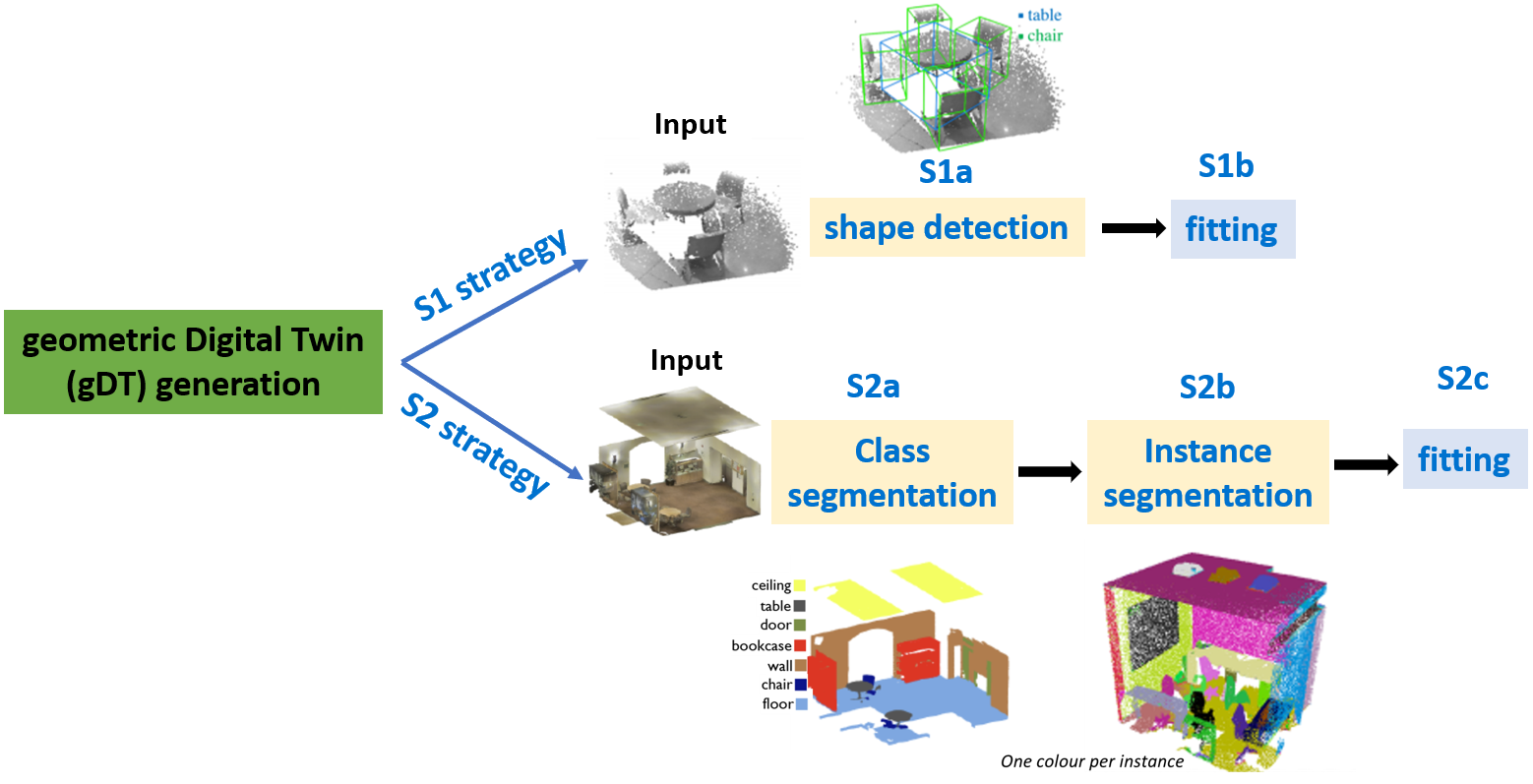}
\caption{Automated geometric Digital Twinning strategies}
\label{fig:CLOI_literature}
\end{figure}

There are two distinct gDT generation strategies investigated in the literature as presented in Figure \ref{fig:CLOI_literature}. The first one ({\bf S1}) involves two steps: (a) primitive industrial shape detection and (b) fitting. The second one ({\bf S2}) has three steps: (a) class segmentation, (b) instance segmentation and (c) fitting. Class segmentation describes the procedure of partitioning the TLS point cloud dataset to clusters of points with class labels assigned per point (such as cylinder, elbow, I-beam, valve, flange, angle and channel) \cite{Li2019ASegmentation}. Instance segmentation assigns a label per point based on the individual object that the point belongs to. For reasons explained in \cite{agapaki2020cloi}, the {\bf S2} DT strategy was selected in this paper. Therefore, the literature review is elaborating on: (a) {\bf S2} class segmentation methods and (b) {\bf S2} instance segmentation methods. Fitting methods are not discussed, since they are out of scope of this paper.

\subsection{Class segmentation}

Class segmentation methods applied on industrial shapes have been widely investigated. We categorize them into three groups: (a) attribute based methods, (b) machine learning and (c) deep learning methods. A comprehensive review of class segmentation methods based on hand-crafted features is provided by \cite{Agapaki2020ChapterGeneration} and some of the most important methods are explained in the paragraphs that follow.

\paragraph{Attribute-based} Attribute-based methods are bottom-up approaches that cluster base elements to generate complex systems in successive higher levels until a top-level system is formed (e.g. bridge, facility) \cite{Borenstein2008CombinedSegmentation}. These methods cluster points with similar attributes into subsets. An $n$-dimensional attribute space is created to extract the attributes in the parameter domain, where $n$ represents the estimated number of attributes. These methods process a point cloud starting from point-wise features and generate higher-level features, such as surface normals \cite{Rusu2009Close-rangeEnvironments,Sampath2010SegmentationClouds}, mesh \cite{Marton2009OnClouds} or patches \cite{Vosselman2009AdvancedProcessing,Zhang20153DSensor}. The estimated attributes are clustered and extracted in the parameter domain. Attribute based methods can be divided in two broad categories based on the shape descriptors they use: global or local. Local descriptors allow for partial matching of features, therefore are preferred for occluded scenes compared to global descriptors. Global descriptors describe the scene as a whole. For instance, local descriptors of a cylinder are curvature and normal vectors, whereas global descriptors are its length and diameter, which correspond to properties for the whole cylinder. Curvature has been extensively used as a local feature for industrial piping segmentation \cite{Dimitrov2015SegmentationSystems,Perez-Perez2016SemanticSegmentation}. However, substantial manual segmentation is needed to pre-process the input TLS data. 

\paragraph{Machine learning} We review one of the most widely used parametric supervised machine learning methods in the class segmentation literature, which is {\bf Support Vector Machines (SVMs)}. \cite{7478077} used SVMs on TLS urban point clouds and then a multi-classification graph-cut algorithm to optimize the initial segmentation result. Similarly, \cite{Zhang_2013} used a region-growing algorithm before applying an SVM for urban point cloud segmentation. \cite{Huang2013DetectingApproach} and \cite{Armeni20163DSpaces} use SVM classifiers with local features to segment cylindrical and indoor space objects. The use of SVMs in these approaches though has inherently two limitations: (1) SVM is not designed for imbalanced classes. Weights inversely proportional to the class frequency are applied to the imbalanced classes. Industrial facility datasets are highly imbalanced with respect to the most important object types they have, since their distribution follows the Zipf's law as proved in \cite{Agapaki2018PrioritizingFacilities}. For this reason, the application of SVMs on TLS industrial facility data is not preferred, unless one oversamples the object types that appear less frequently. (2) the success of SVMs depends on the selection of hand-crafted features, the type of kernel function and the parameters to the kernel function. Improper selection of features can result in misclassifications, whereas application of different kernel functions for a dataset gives different results.

\paragraph{3D Class Segmentation Deep Learning methods} CNNs have been widely used for a variety of tasks in image segmentation \cite{Krizhevsky2012ImageNetNetworks,LeCun2008BackpropagationRecognition,Taha2015MetricsTool,Pang2012ComputerizedAlgorithm,Wang2018DeepNetworks,Teichmann2018MultiNet:Driving}. We group these methods in three main categories as suggested by \cite{Wang2019AssociativelyClouds}: ({\bf DLa}) view-based \cite{Su2015Multi-viewRecognition,Kalogerakis20173DNetworks,Wei2016DenseNetworks}, ({\bf DLb}) volumetric \cite{Maturana2015Voxnet:Recognition,Wu20153DShapes,Zhou2017VoxelNet:Detection,Klokov2017EscapeModels,Tatarchenko2017OctreeOutputs} and ({\bf DLc}) geometric deep learning methods \cite{Qi2017PointNET:Segmentation,Qi2017PointNet++:Space,Wang2019AssociativelyClouds}. 

Geometric deep learning methods are chosen as the most suitable for class segmentation as explained by \cite{agapaki2020cloi}, since they address the following challenges that TLS industrial point cloud processing has: (1) irregularity in the TLS data structure, (2) TLS data sparsity, noise, presence of outliers and occlusions as well as density variations especially in industrial settings and (3) differences in industrial object scales, rotation and translation variant objects as well as geometric similarities between objects of the same class. PointNETs \cite{Qi2017PointNET:Segmentation,Qi2017PointNet++:Space} and their derivatives \cite{Wang2019AssociativelyClouds,Wang2018SGPN:Segmentation,landrieu2018largescale,thomas2019kpconv} have solved these challenges by applying permutation invariant functions as well as local 3D filters in their network architectures. PointNET networks concatenate global and local features into point feature vectors based on which class labels are predicted. PointNET++ improves the PointNET architecture by adding local neighbourhood geometric features.

\subsection{Instance Segmentation}  

3D instance segmentation is based on 3D geometric class segmentation networks. These methods can be grouped into shape-based (top-down) or shape-free (bottom-up). Our readers can refer to \cite{agapaki2020instance} for a comprehensive literature review of each of these methods. We elaborate on the state-of-the-art literature on shape-free methods, since these are more suitable for the generation of gDTs from TLS industrial data \cite{agapaki2020instance}.

Shape-free methods are based on deep learning networks, which aggregate features per point and output instance labels per point given a similarity matrix between pairs of points \cite{Wang2018SGPN:Segmentation,Wang2019AssociativelyClouds} or embedding another network measuring point-wise distances \cite{Pham2019JSIS3D:Fields}. PointNET \cite{Qi2017PointNET:Segmentation} and PointNET++ \cite{Qi2017PointNet++:Space} is the backbone network for these methods, meaning that they achieve class segmentation as well. Although these networks take into consideration the local neighbourhoods of points, they cannot explicitly define the boundaries of complex industrial shapes. Object boundaries can be taken into account by considering the class and instance segmentation labels. The readers can refer to \cite{xie2019review} for a detailed review of all the instance segmentation methods.

\section{Proposed solution}
\label{proposed}

We target to solve the problem of the generation of gDTs of existing industrial facilities with respect to cost and modelling time reduction. The main objective of this paper is to develop a benchmark framework as the foundation for future research. 

\subsection{Overview}

The proposed framework consists of two major parts. Specifically, these parts are {\bf (1) class segmentation} and {\bf (2) instance level segmentation} that intend to answer the research questions as outlined in the Background section and aim to outperform the existing state of practice and research in the industrial modelling space. 

We propose a novel hybrid framework which develops deep learning networks and leverages their detected outputs with industrial engineering knowledge, in order to automatically extract labelled point clusters corresponding to industrial shape components without generating surface primitives  ({\bf class point clusters}) and then to efficiently detect individual industrial shapes from the labelled point clusters ({\bf instance point clusters}).

Real-world industrial environments are more challenging than buildings that have been extensively studied and scanned in previous research efforts as mentioned in the Background section. Industrial components do not comply with a universal colour scheme, rather colours depend on each manufacturer's specifications \cite{agapaki2020cloi}. Industrial spaces are typically large and unstructured with shapes that may span across their whole length/width and they are heterogeneous spaces where there are usually no direct contextual rules in separate systems (piping, structural, electrical) and only the components that belong to the same system are internally connected with strong context. For example, the relative location of a cylinder with respect to an I-beam in a factory does not imply that the locations of these objects should comply to specific spatial rules. We propose a 3D-slicing facility window method, CLOI-NET-class based networks and CLOI-Instance graph-connectivity algorithms to tackle these challenges. The 3D windows are used to segment the TLS dataset in non-overlapping parts, so that a portion of these windows will be used for training. These windows should be non-overlapping, so that the training and test set are disjoint. These algorithms are the core foundation of the methods built upon them to enhance the segmentation and detection results. The proposed algorithms can deal with the challenges outlined above and can accurately detect the majority of \textit{CLOI} industrial shapes.

Most of the \textit{CLOI} shapes match 1 to 1 to a component class, (i.e. the shape is unique to this component), but for cylinders the shape is not unique. So the method focuses on {\bf segmenting the \textit{CLOI} shapes}, and by default, equivalently segments their component classes except for cylinders. Segmentation of the subcategories of cylindrical shapes (i.e. pipes, circular hollow sections, handrails, electrical conduit) is beyond the scope of this research. The proposed framework is not applicable for connections of steel members (welding and bolting). The proposed algorithms address scale variance (The algorithms are scale invariant, since we feed them with objects at different scales (from a few centimeters to some meters.) of industrial objects and intra-class variations. For instance, there are many types of valves as expressed above, which are grouped in one class and the proposed algorithms should be able to segment valves of all the above mentioned categories.

\begin{figure}[!ht]
\centering
\includegraphics[width=0.9\textwidth]{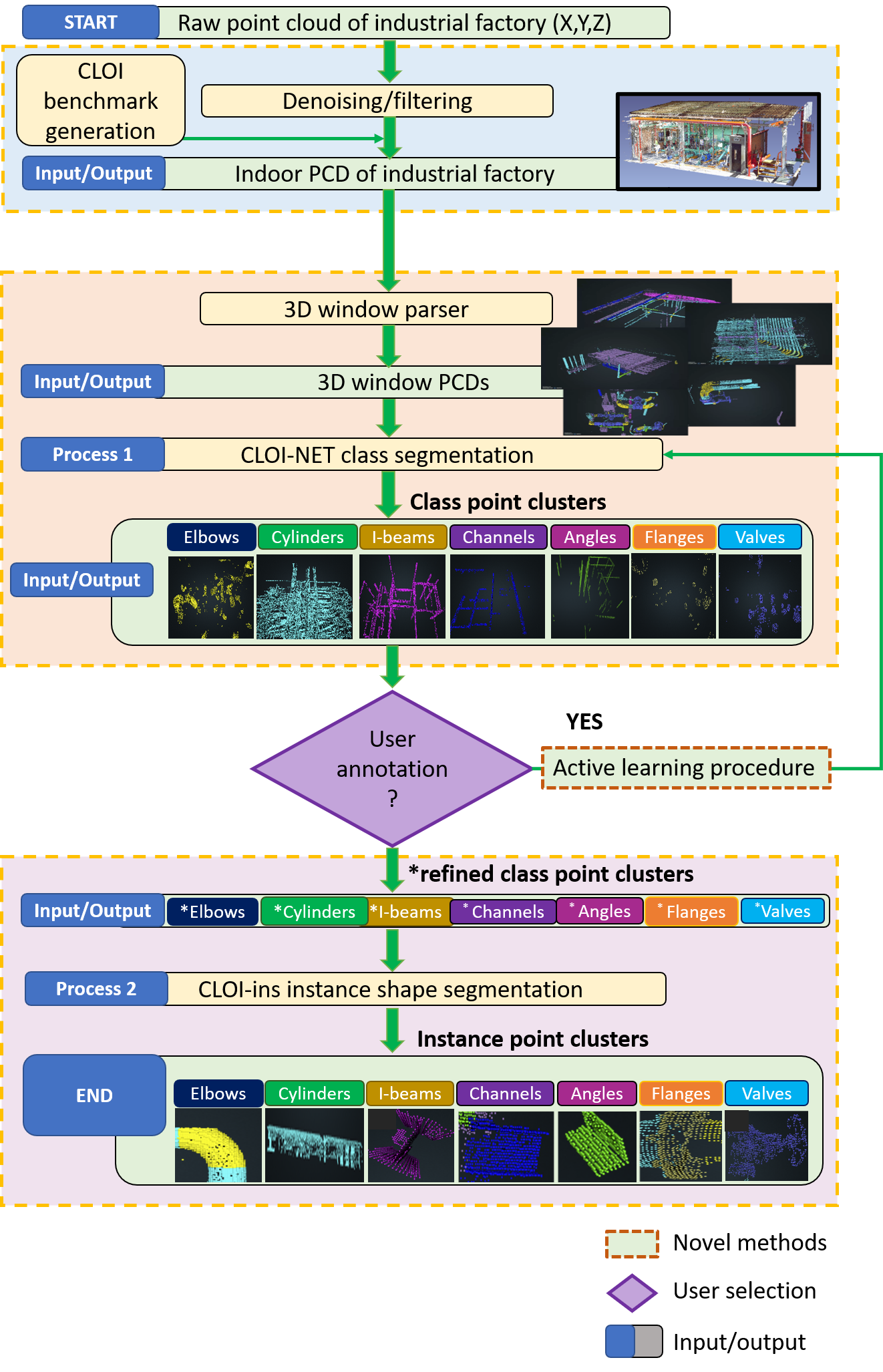}
\caption{Workflow of the proposed \textit{CLOI} framework}
\label{fig:framework}
\end{figure}

We illustrate the developed hybrid framework in Figure \ref{fig:framework}. It consists of two major processes: {\bf Process 1}, class segmentation of \textit{CLOI} industrial point clusters, and {\bf Process 2}, instance segmentation of \textit{CLOI} industrial shapes from point clusters.

The proposed framework starts with a raw, laser-scanned, PCD of an existing industrial facility (data format: points in .pcd, .txt, .las, .xyz). External noise such as vegetation, adjacent buildings is removed using commercial software as explained in \cite{agapaki2020cloi}. The industrial PCD contains \textit{CLOI}-shapes and any other industrial shapes inside a factory (data format: points in .pcd, .txt, .las, .xyz). The first step of the framework is to automatically split the PCD facility in 3D windows and the 3D windows in ``3D blocks''. Then, the 3D blocks are aligned in the global coordinate system. As such, the outputs of this step are 3D block PCDs (data format: points in .pcd, .txt, .las or .xyz). Then, we manually annotate industrial facilities to generate a benchmark dataset and the outputs of this step are class and instance segmentation labels and points. It is important to note that this is an essential offline step needed for training purposes and serves as the ground truth for the validation of the framework. 

Next, we propose a three-step class segmentation method (Process 1) to segment the \textit{CLOI} point clusters from the 3D blocks. The final outputs of this process are seven industrial shapes, namely cylinders, elbows, channels, I-beams, angles, flanges and valves, in the form of labelled point clusters (data format: points in .pcd, .txt, .las, .xyz). Then, we suggest an optimal manual annotation (if the users select it) to remove the erroneous point clusters maintained from Process 1 followed by proposing an efficient instance segmentation method (Process 2) through which the seven \textit{CLOI} classes (in point cluster format) can be directly segmented to individual shapes. The final outputs of this process are point data corresponding to the points, class and instance labels per point. We elaborate on each process in the following sections.

\begin{figure}[!ht]
\centering
\includegraphics[width=\textwidth]{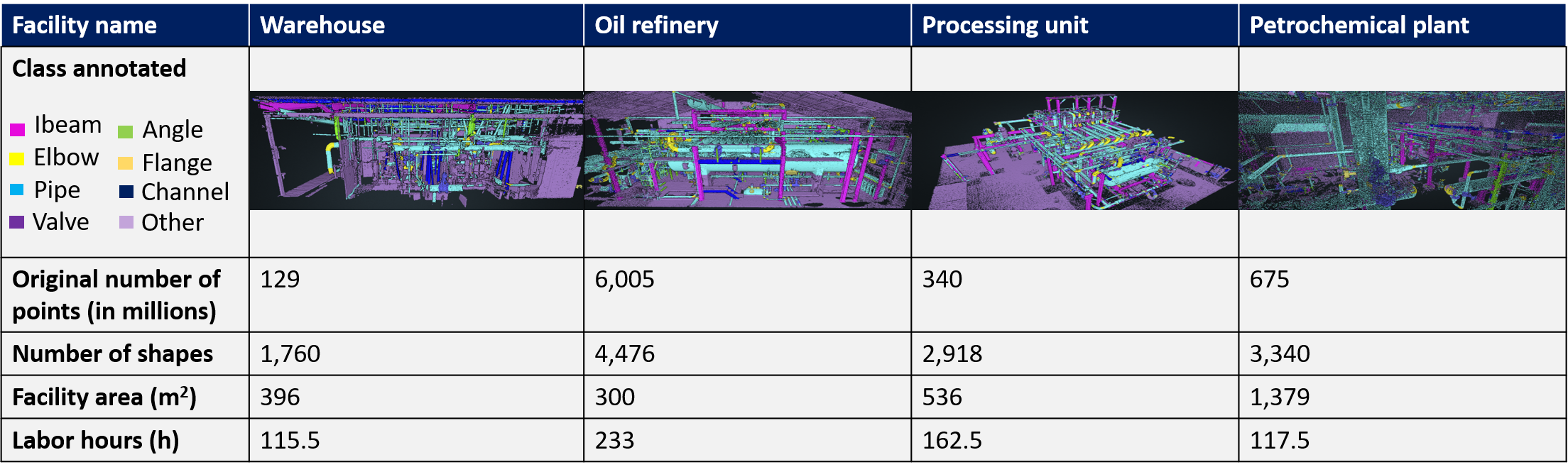}
\caption{\textit{CLOI} benchmark dataset specifications}
\label{fig:CLOI_benchmark}
\end{figure}

We validate Process 1 on the \textit{CLOI} benchmark dataset \cite{Agapaki2019CLOI:Facilities}, which is composed of four laser scanned industrial facilities. The original number of laser scanned points, the number of instances, the area and the manual labor hours to manually annotate (with class and instance labels) each facility are documented in Figure \ref{fig:CLOI_benchmark}.

\subsection{Process 1: CLOI-NET-Class segmentation} 

The methods of Process 1 bypass the stage of surface generation altogether and directly output segmented and labelled point clusters. The 3D window parsing method breaks down the whole industrial facility into subset windows for more efficient processing. The key insight behind Process 1 is to formulate a high dimensional feature space to automatically assign labels per point so that the target point clusters can be quickly located in the point cloud.

The inputs of the method are the spatial coordinates of TLS points and the outputs are labelled, segmented point clusters with confidence levels of the predictions. Here we define segmented point clusters as all the points that belong to one class i.e. all cylinder points is one class point cluster. The method consists of three major steps: {\bf Step 1} partitions each facility into smaller spaces using a 3D sliding window/block approach and prepares the data for training, {\bf Step 2} predicts a class label per point using a modified version (SFR) of a geometric deep learning network for point cloud segmentation (PointNET++) with the goal to accurately segment the \textit{CLOI} shapes. In {\bf Step 2}, the user has two options on how to train the network, either training with no data from the test facility or manually annotating data of the test facility and including those for training. The latter is based on the assumption that, inevitably, any class segmentation algorithm will have errors, which will have to be manually corrected eventually. Therefore the goal is to minimize the total manual annotation time. {\bf Step 3} refines the predicted class labels by improving class level predictions with stronger contextual relationships. 

The success of the proposed pipeline is measured not by maximising the point-wise accuracy of the method, rather by minimising the cost that it incurs to the modelers when using it. This novel method leverages the advances in point cloud deep learning segmentation, contextual shape specific attributes and active learning in order to accurately predict point-wise class labels with no significant difference in performance for diverse industrial environments. A critical part of this method's novel design is the stage-wise annotation, which permits both human-annotated and automatically annotated points to influence the system’s view of what needs the most human attention next. Details of our methodology, named \textit{CLOI-NET-Class}, can be found in \cite{agapaki2020cloi}.

\subsection{Process 2: CLOI-Ins instance segmentation}

The inputs of Process 2 are the predicted point clusters from the CLOI-NET-Class method for the evaluation of the proposed framework. The same 3D block generation method from Process 1 is used for segmenting the input data. The outputs of this process are point-wise instance labels (individual point clusters of \textit{CLOI} shapes).

Process 2 consists of two major steps: Step 1 predicts an instance label per point by using a graph-based method, namely Breadth First Search (BFS) that was originally introduced by \cite{Bauer1972TheLanguages}. Step 2 is a boundary segmentation method that is used to enhance the instance segmentation results of Step 1. An assumption of the method is that the initial TLS industrial data is partitioned in 3D non-overlapping sliding windows with overlapping 3D blocks. The outputs of Step 1 are connected components based on connectivity relationships in order to segment the instances as output. The boundary segmentation method in Step 2 outputs binary labels on whether a point is a boundary point or not. These instance point clusters present industrial shapes at Level of Detail (LOD) 300. 

The novelty of Process 2 is two-fold: 

\begin{enumerate}\setcounter{enumi}{0}
    \item the efficiency of the BFS algorithm by applying it on the entire PCD and connectivity between points
    \item the intelligence of the boundary segmentation method to account for boundary points and robustly process points in small regions.
\end{enumerate}

Readers can refer to \cite{agapaki2020instance} for details of the CLOI-Ins instance segmentation process.

\section{EXPERIMENTS AND RESULTS}
\label{methodology}

\subsection{Implementation}
\label{researchActivities}

The author generated the first dataset of class labelled point clusters of industrial facilities, \textit{CLOI}, \cite{Agapaki2019CLOI:Facilities} to validate Processes 1 and 2. \textit{CLOI} consists of 10 classes that cover a wide range of industrial scenes (both indoor and outdoor). The TLS datasets of four laser scanned industrial facilities are used for the generation of \textit{CLOI} as shown in Figure \ref{fig:CLOI_benchmark}. One facility is a warehouse, one is a petrochemical plant, one an oil refinery and the fourth a processing unit. These facilities are anonymized since rights are reserved by AVEVA Group Plc. and British Petroleum. All datasets were obtained using static terrestrial laser scanners. This research provides the (to the best of our knowledge) hitherto largest collection of terrestrial laser scans of industrial facilities with point-level (a) class and (b) instance ground truth annotations. (a) refers to one of the ten \textit{CLOI} classes and (b) is an index number that refers to a specific individual shape and is not further used in this work. In total, it consists of 12,497 shapes and 7.1 billion points with their class and instance labels for each point. To this end, this research provides \textit{CLOI}, the largest annotated dataset based on already existing datasets \cite{agapaki2020cloi} and the only dataset of industrial environments that is captured with more than one sensors. This means that processing \textit{CLOI} point cloud data is independent of the data capturing system that was used to generate the data. \textit{CLOI} is also the only dataset available for processing PCDs of industrial environments. Detailed statistics and scanner specifications of the data can be found in \cite{Agapaki2019CLOI:Facilities}. 

Two research platforms were developed for the framework validation; one capable of high computing for training deep neural networks and one for visualisations of large scale TLS industrial datasets. Training of the CLOI-NET-Class method was performed on Google Cloud instances. We implemented the deep learning class segmentation experiments on Tensorflow 2.0 as a proof of concept prototype and ran experiments on Google Cloud (Deep Learning VM image) with NVIDIA Tesla P100 GPUs. Visualizations of point clouds and segmentation results were implemented on the \textit{CLOI} platform which is based on the Potree Viewer (http://potree.org/) in JavaScript. Potree is built upon ThreeJS and allows for rendering of large point clouds in a WebGL web browser \cite{Schuetz2016Potree:Browsers,Devaux2012AModels}. We created the user interface to select the TLS dataset of a \textit{CLOI} facility, then segment the \textit{CLOI} classes and validate with the ground truth class labels. The user can also select a point and only view the points associated with that \textit{CLOI} class. Further details about the implementation of Process 1 and Process 2 can be found in \cite{agapaki2020cloi} and \cite{agapaki2020instance} respectively.

\subsection{Manual annotation}

The \textit{CLOI} dataset was generated by manually annotating the four industrial facilities. The Ground Truth (GT) datasets are the desired outputs to compare against those generated by the proposed methodology and also used for training. The following GT datasets were created for the \textit{CLOI} dataset validation.

{\bf GT class:} A given industrial facility, TLS scanned, point cloud input is segmented into the eight \textit{CLOI} classes. Each individual point was assigned a class point-wise label. Figure \ref{fig:CLOI_benchmark} shows each \textit{CLOI} facility coloured with one of the eight class labels and the manual annotation time involved to generate the GT per facility. The number of shapes (instances), original number of 3D points and the area per facility are also provided. One can distinguish that even if a small facility area is scanned, the density of the scans may be so high that the number of points is much higher compared to a sparsely scanned facility. For instance, the \textit{oil refinery} is only $300m^2$, making it the smallest facility of the dataset, but it has the largest number of surveyed 3D points. 

{\bf GT instance:} A given point cloud input is assigned to an individual instance point cluster. 

{\bf GT boundary:} A given point is classified as a ``boundary'' point if there is more than one instance in a neighbourhood of radius $4cm$ around it. The data structure used to define the neighbourhoods around each point is a kDTree. 

\subsection{Experiments}

The performance of the framework was evaluated based on: 

\begin{enumerate}\setcounter{enumi}{0}
  \item the performance of the \textit{CLOI-NET-Class} segmentation network 
  \item the performance of the \textit{CLOI-Instance} segmentation network.
\end{enumerate}

The inputs of the proposed framework are the class segmented clusters of Process 1. The class segmentation experiments showed average accuracy and mIoU of 79.8\% and 44.65\% when all the \textit{CLOI} facilities are included for training except the one of interest to segment that is tested. CLOI-NET-Class has been proven to be consistent, reliable and without significant bias when tested on all the \textit{CLOI} facilities. The author validated the theoretical active learning model as outlined in \cite{agapaki2020cloi}. Results showed that the total cost annotation function and the validation accuracy follow the theoretical model and the optimal data pre-annotation percentage that minimized the total annotation cost is between 20$\pm$10\%. The CLOI-NET-Class performance following the active learning approach had on average 15\% higher accuracy than the passive learning approach.

\begin{figure}[!ht]
\centering
\includegraphics[width=\textwidth]{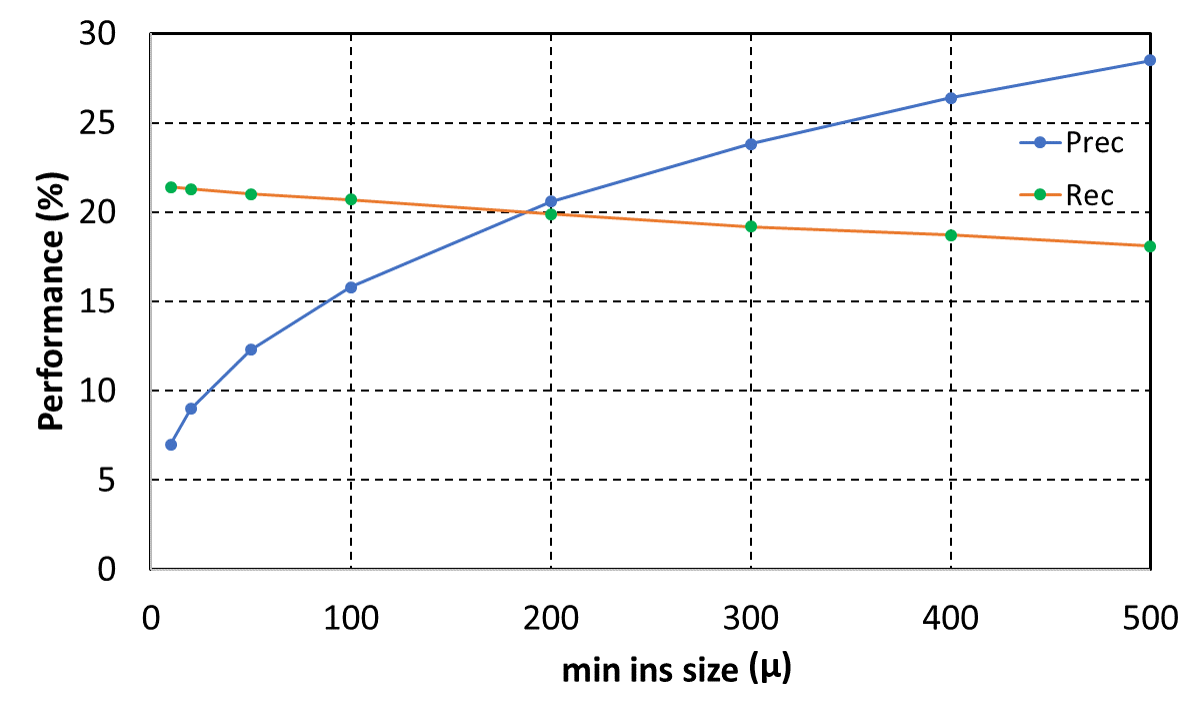}
\caption{Performance of the BFS algorithm with respect to the minimum instance size ($\mu$) for IoU=50\% and $\epsilon=4$ cm. Test on the oil refinery facility.}
\label{fig:mininsSize_pred}
\end{figure}

The performance of Process 2 (CLOI-Ins segmentation) was 73\% mPrec and 71\% mRec on all \textit{CLOI} facilities using the ground truth class labels as inputs \cite{agapaki2020instance}. For the evaluation of the framework, we compared the state-of-the-art instance segmentation networks (SGPN \cite{Wang2018SGPN:Segmentation,Wang2019AssociativelyClouds}), the BFS algorithm and the proposed \textit{CLOI} framework in Table \ref{table:inscomparisonexp}. The results illustrated in Table \ref{table:inscomparisonexp} show that SGPN has very low performance on the oil refinery data with the ASIS network performing better in all efficiency metrics. The oil refinery is used to compare the state-of-the-art deep learning instance segmentation networks, the BFS algorithm and the \textit{CLOI} framework methodology. For the application of the BFS algorithm, the minimum instance size was selected for the predicted \textit{CLOI} class point clusters based on performance. Therefore, the author conducted experiments to determine the minimum instance size based on the performance in terms of precision and recall on the \textit{CLOI} datasets. The results in Figure \ref{fig:mininsSize_pred} illustrate that the optimal trade-off between precision and recall is for minimum instance size 200 points instead of the minimum instance size of 20 points that was computed based on the ground truth class segmentation labels \cite{agapaki2020instance}. This is attributed to noisy predicted class labels compared to the ground truth class labels used to evaluate Process 2 independently. There is an exception for the minimum instance size ($\mu$) and the minimum neighbourhood size ($\epsilon$) for the case of cylinders. The results indicate to set the instance size at 50 points and the minimum neighbourhood size ($\epsilon$) at 3cm (instead of 4cm) only for cylinder instance point clusters due to the observation that cylinders have higher class segmentation label predictions and the CLOI-Instance methodology benefits from that. We also observe in Table \ref{table:inscomparisonexp} a 10\% increase in precision due to the \textit{class boundary} constraint on the BFS algorithm for a minimum neighbourhood of 4cm. 

The author then tested the performance of the same methods per \textit{CLOI} shape in Table \ref{table:ASISSGPNexp}. We present these results for the oil refinery dataset as an example for comparison of the best performing existing instance segmentation methods and the proposed \textit{CLOI} framework. The illustrated results in Table \ref{table:inscomparisonexp} and \ref{table:ASISSGPNexp} demonstrate that the CLOI-Instance methodology clearly outperforms the current state-of-the-art research.

Another important note is that the CLOI framework results are calculated assuming that the users pre-annotate X\% of the test facility with X\% being the value from Table \ref{table:activeX} depending on the facility. These percentages are based on the active learning curves of \cite{agapaki2020cloi}. 

Then, we present the precision and recall per \textit{CLOI} class and the average precision and recall curves in Figure \ref{fig:BFS_BP_pred} as a reference. The results for the other three facilities are included in the Appendix. It is evident that for all datasets the recall metric of all the \textit{CLOI} classes outperforms the precision metric for all the IoU threshold values. The greater difference between the mean precision and mean recall is for the oil refinery (Figure \ref{fig:BFS_BP_pred}(c)), which is attributed to the high complexity of this dataset. This leads to reduced performance for all classes. Although the CLOI-Instance proposed methodology has promising results compared to the state-of-the-art methods for the instance segmentation task, the results demonstrate that the predicted class labels significantly reduce the precision and recall metrics compared to the same results presented given the ground truth class labels \cite{agapaki2020instance}.

The CLOI framework performance of cylinders is relatively high across the \textit{CLOI} facilities given their high class segmentation performance \cite{agapaki2020cloi} for all the IoU threshold values. We remind the reader that the cylinder class segmentation performance was 81.25\% precision, 81.75\% recall and 68.25\% IoU on average. There are though some cases where the cylinder instance point clusters are over- or under-segmented. These cases are the {\bf Cyl} cases presented in \cite{agapaki2020instance}. The results of the \textit{CLOI} framework show an additional pain point. This is the uncertainty of the CLOI-NET-Class segmentation on predicting the class labels of the points. This leads to erroneous instance label predictions and mostly impacts the \textit{CLOI} classes that have low class segmentation performance (the reader can refer to \cite{agapaki2020cloi} for a detailed discussion). 

\begin{figure}[!ht]
\centering
\includegraphics[width=\textwidth]{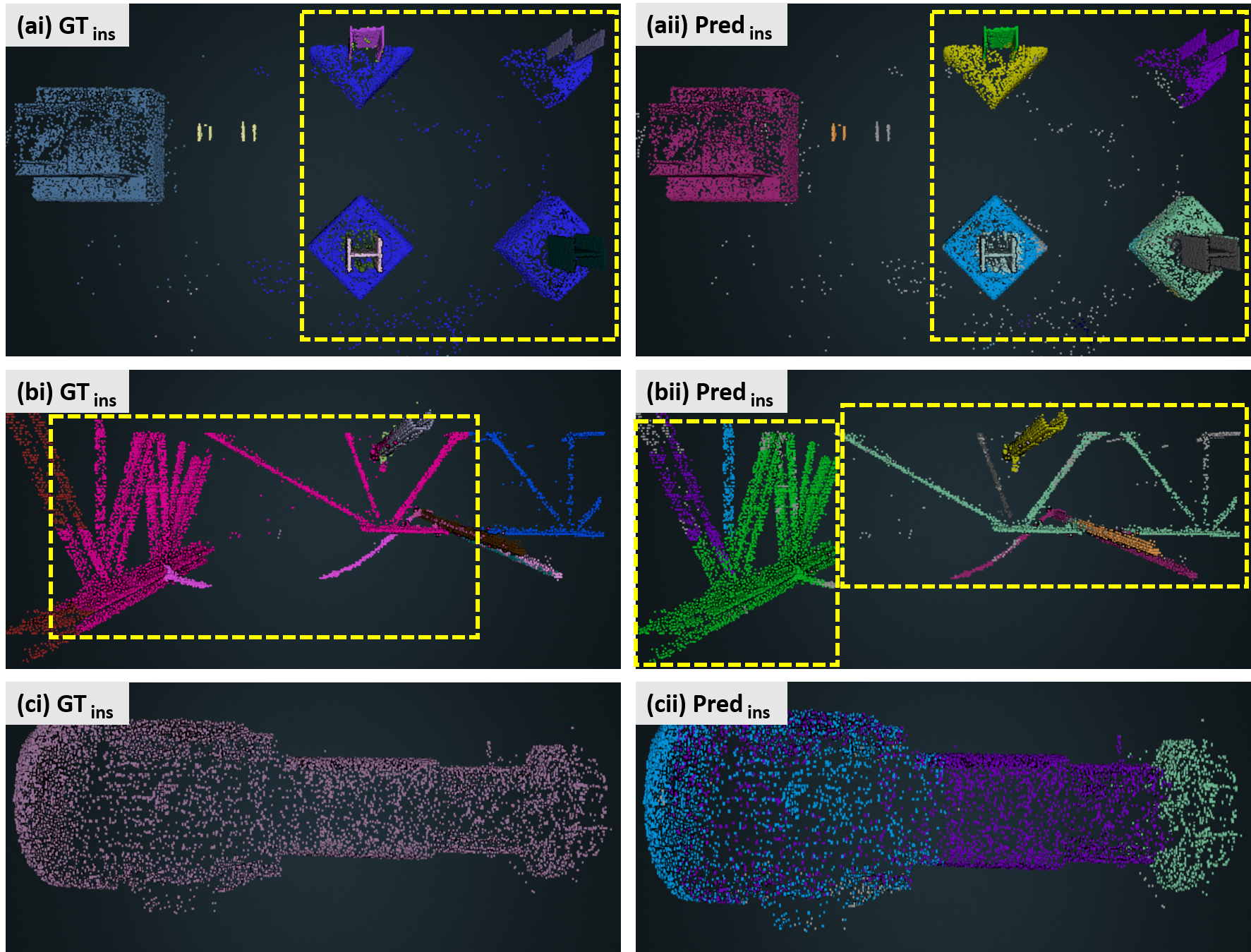}
\caption{Examples where the CLOI framework outperforms the manual instance segmentation. (i) refers to ground truth instances and (ii) refers to predicted instances with the CLOI framework.}
\label{fig:CLOIframe_outperform}
\end{figure}

Another achievement of the \textit{CLOI} framework is that it correctly segments sub-instances of an instance point cluster that has the ``other'' class label and even outperforms the manual instance segmentation in cases where a ground truth instance is under-segmented (Figure \ref{fig:CLOIframe_outperform}(a) and Figure \ref{fig:CLOIframe_outperform}(b)). This particularly applies for instances close to the floor or roof of a facility. The superior performance of the \textit{CLOI} framework is attributed to the connectivity information that the BFS algorithm uses to segment instances. Another case where the \textit{CLOI} framework outperforms the manual instance segmentation is for sequences of pipe components that have different radii. An example of that is Figure \ref{fig:CLOIframe_outperform}(c) where the \textit{CLOI} framework correctly segmented the cylinder from a pump and a flange with steel rods.

We then recommend to use the 25\% IoU threshold that gives slightly improved results (50\% mPrec and 35.3\% mRec for all the \textit{CLOI} facilities). The \textit{CLOI} shapes that have significantly higher metrics are those with higher class segmentation results as explained above. These are cylinders (53.6\% mPrec and 44\% mRec), elbows (66.8\% mPrec) and I-beams (63\% mPrec and 64.3\% mRec). 

\subsection{Time savings in Geometric Digital Twinning}

\begin{figure}[!ht]
\centering
\includegraphics[width=\textwidth]{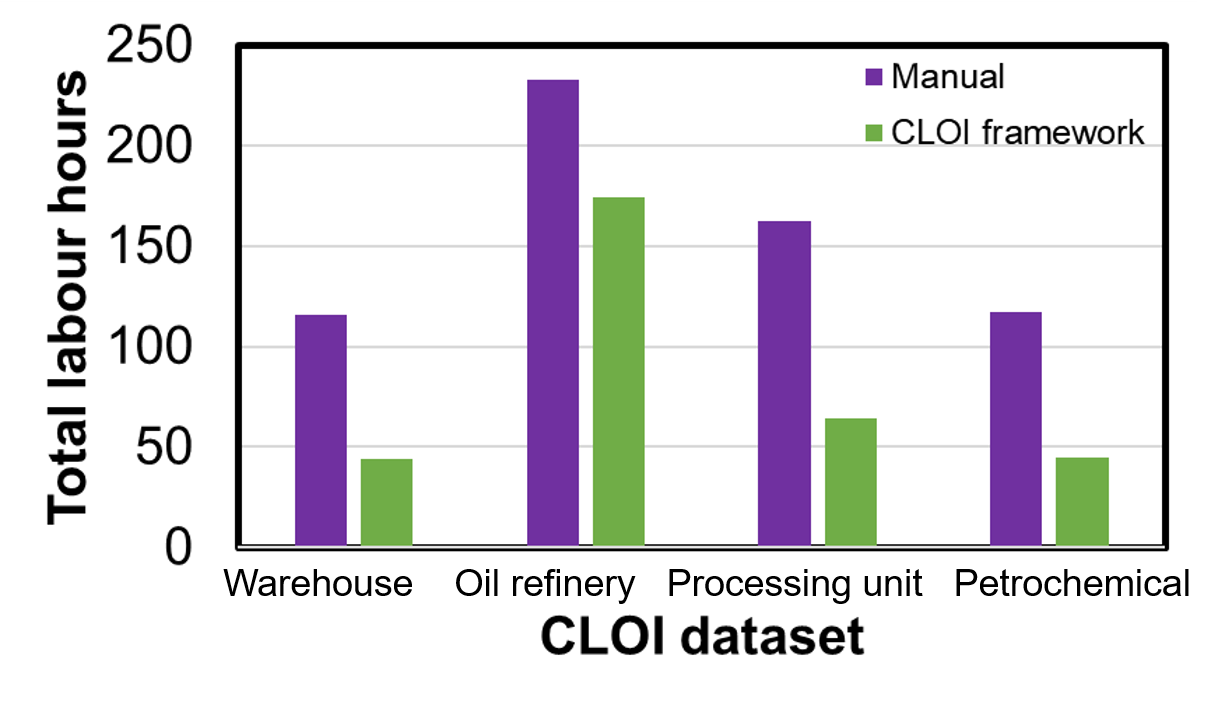}
\caption{Manual and our framework's total labour hours per \textit{CLOI} dataset.}
\label{fig:TotalSegmentationSavings}
\end{figure}

One of the main goals of Process 2 was to prove that the CLOI-Instance method requires competitively less manual segmentation time compared to the current practice. We validated this hypothesis for the overall framework given that the class segmentation labels are predicted from the CLOI-NET-Class method (Process 1). We use the percentage of \textit{CLOI} shapes that the CLOI-Instance method correctly predicts as a proxy to approximate the number of manual labour hours that are still needed in order to achieve an accurate gDT generation. The results are summarized for each \textit{CLOI} dataset in Table \ref{table:savingsInstance}. A comparison of the manual instance segmentation time for the \textit{CLOI} benchmark dataset generation and the \textit{CLOI} overall framework segmentation time is presented in Figure \ref{fig:TotalSegmentationSavings}. The total number of man hours needed when deploying the overall framework is calculated as follows. The number of manually segmented \textit{CLOI} shapes is computed as the product of the number of shapes that are missed by the framework ($1-recall$) and the average time it takes a modeller to manually segment a given shape. An assumption for the simplification of the calculation here that each \textit{CLOI} shape takes the same time regardless of its complexity. The results illustrate that 35\% of the manual labour hours are saved on average. The oil refinery dataset is one of the most complex \textit{CLOI} datasets and this is reflected in reduced savings in labour hours for instance segmentation. It is noteworthy that for all \textit{CLOI} facilities, the cylinder \textit{CLOI} shapes have relatively low recall ($\approx 40\%$) which is attributed to the large number of conduit that are clustered together in one instance. 

We evaluated in \cite{Agapaki2018PrioritizingFacilities} the state-of-the-art commercial software that semi-automatically segments cylinders from TLS industrial datasets, however a direct comparison cannot be made since the total number of cylinders considered in that evaluation does not match the number of cylinders in the \textit{CLOI} dataset. However, the number of cylinders correctly detected by EdgeWise can be compared with the number of cylinders segmented by the proposed framework. The results in Table \ref{table:cylindersComparison} demonstrate that the proposed framework correctly segments more cylinders than those detected by EdgeWise. The proposed framework is designed to better segment conduits and even with the discussed limitations, Table \ref{table:cylindersComparison} illustrates its superiority to EdgeWise which is mostly in the correctly predicted conduits that EdgeWise does not identify.

\begin{figure}[!ht]
\centering
\includegraphics[width=\textwidth]{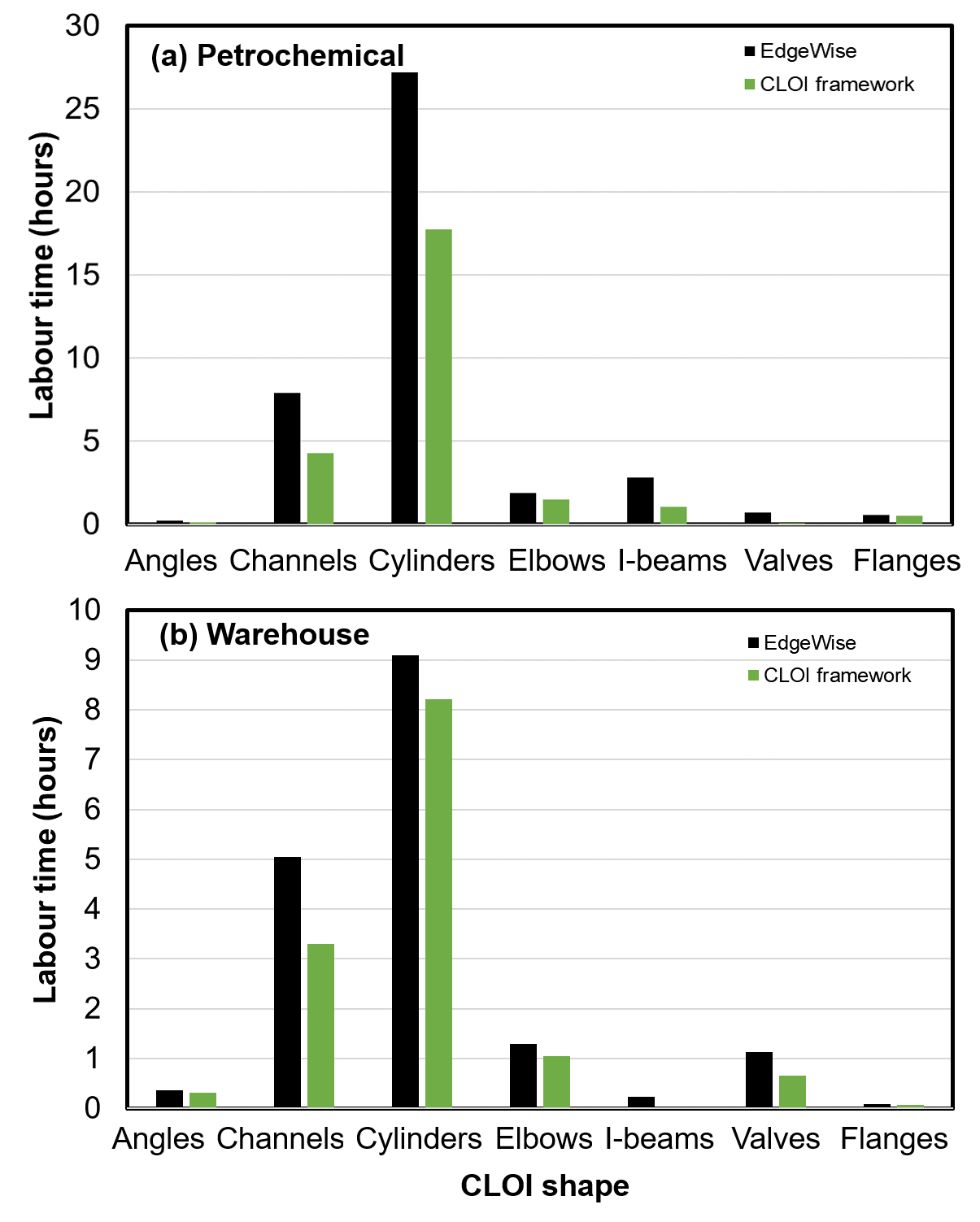}
\caption{Comparison of EdgeWise and our framework with respect to manual labour hours per \textit{CLOI} shape for the (a) petrochemical and (b) the warehouse dataset.}
\label{fig:SavingsEdgeWise}
\end{figure}

The performance of the proposed framework is then compared directly with EdgeWise assuming that the modeling of \textit{CLOI} shapes will be manually performed in EdgeWise. Therefore, the average modeling labour time per object is taken from \cite{Agapaki2018PrioritizingFacilities} and multiplied with the number of objects that are not automatically segmented. The output in labour hours in shown in Figure \ref{fig:SavingsEdgeWise} and compared with the manual labour hours for the objects that EdgeWise cannot automatically detect (a fraction of cylinders and the rest of \textit{CLOI} shapes). Figure \ref{fig:SavingsEdgeWise} shows that 21\% and 39\% more time savings are achieved when the proposed framework is utilized for the warehouse and petrochemical plant respectively. 

\begin{figure}[!ht]
\centering
\includegraphics[width=\textwidth]{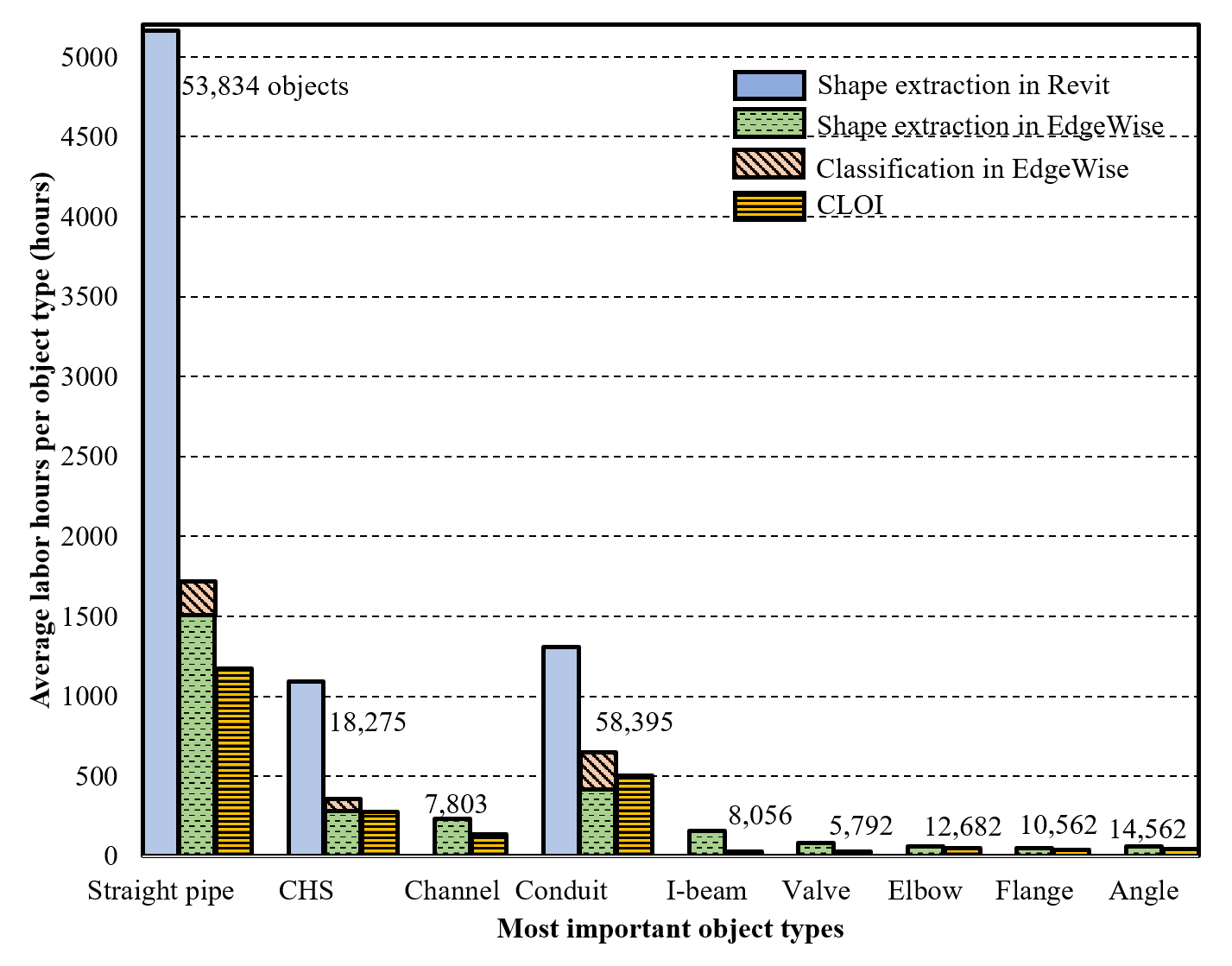}
\caption{Average modelling labour hours per object type for the most important objects of a sample facility with shown numbers of objects.}
\label{fig:SavingsCLOIFinal}
\end{figure}

The warehouse and the petrochemical plant datasets are then used as a proxy to estimate the average percentage of labour hour reduction of the CLOI framework compared to EdgeWise per \textit{CLOI} class. The average percentage per class is shown in Table \ref{table:savingsCLOI}. An assumption was made that the modeling time of all cylindrical shapes is the same, since our framework detects cylinders and not their sub-classes, i.e. pipes. Then, the \textit{CLOI} framework is directly compared with EdgeWise for the petrochemical plant with 240,687 objects that was used for manual modeling in \cite{Agapaki2018PrioritizingFacilities}. The same assumptions are used here for consistency of the results. The results in Figure \ref{fig:SavingsCLOIFinal} reveal that 12 person-months are needed when using the \textit{CLOI} framework instead of the 17 person months that are needed when using EdgeWise. In particular, \textit{CLOI} saves 10\% more man-hours for cylinder modeling, which is translated in 773 labour hours saved. Although there is still time required for manual cylinder extraction, the proposed framework clearly outperforms the commercial software EdgeWise.

\section{Conclusions}
\label{conclusions}

This paper presents \textit{CLOI}, an automated benchmarking framework for generating gDTs of existing industrial facilities from point cloud data. This work focuses on the generation of instance point clusters in a cost-effective approach compared to the current practice. The framework consists of two main processes: the \textit{CLOI-NET-Class} segmentation (Process 1), which generates the ten most important industrial objects in the format of class point clusters and \textit{CLOI-Ins} segmentation (Process 2), which segments the class point clusters into individual point clusters. The \textit{CLOI} framework was experimentally validated on the largest published industrial point cloud dataset, which consists of four TLS industrial point clouds. The consistent results on the \textit{CLOI} dataset demonstrate that the proposed framework can reduce the onerous, repetitive manual work of segmenting industrial shapes and therefore reduce the modelling time of the resulting models. In the following paragraphs, we present the contributions ({\bf Con}) and limitations ({\bf Lim}) of the \textit{CLOI} framework in detail.

\begin{figure}[!ht]
\centering
\includegraphics[width=\textwidth]{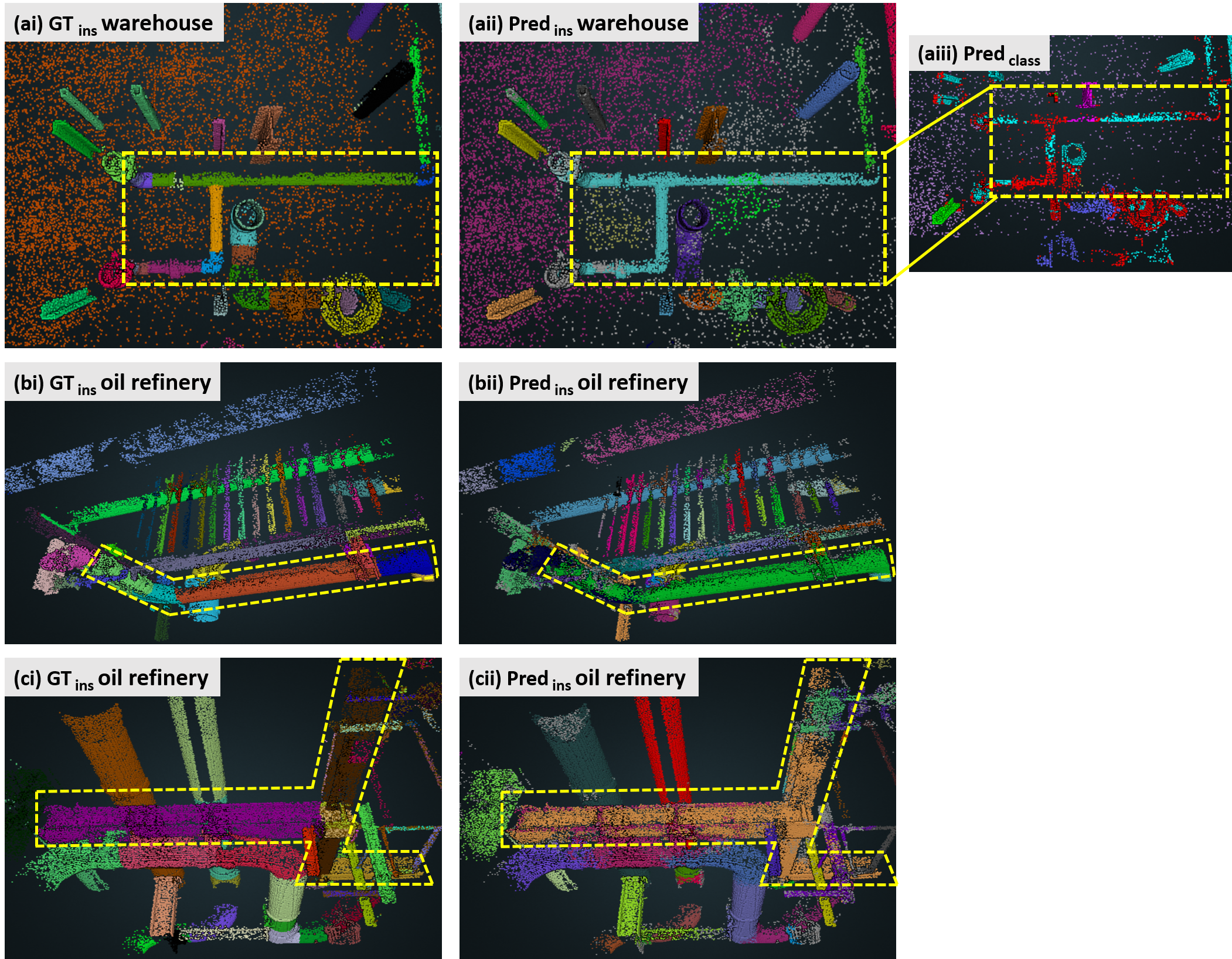}
\caption{Examples of ground truth and predicted instances of piping elements (a,b) and (c) I-beams of a steel frame. (aiii) Predicted class label predictions (predictions with $\leq 80\%$ confidence score coloured in red).}
\label{fig:PipeSpoolFrame}
\end{figure}

{\bf Con 1} This is the first framework of its kind to achieve significantly high and reliable performance (50\% mPrec and 35.3\% mRec) compared to current state-of-the-art research and commercially available software. It is the first framework to provide significant improvements on cylinder segmentation (53.6\% mPrec and 44\% mRec) and the first to segment the rest of the \textit{CLOI} classes. It, therefore, provides a solid foundation for future work in generating DTs of industrial facilities. {\bf Con 2} This research moves forward the state of automated class and instance segmentation from TLS point cloud datasets as well as promotes the value of adding ``intelligence'' to the PCD data. The interpretation of the results strongly suggest that the performance of both the CLOI-NET-Class and the CLOI-Instance methods are significantly improved by using the optimal amount of data during training ($\approx 30\%$) and contextual enforcement rules to accurately segment the \textit{CLOI} classes. {\bf Con 3} It is the first framework of its kind to significantly reduce the manual labour hours (by at least 33\%) compared to the state of practice, EdgeWise. It also has 21\% and 39\% more time savings when segmenting the warehouse and the petrochemical facility dataset compared to EdgeWise. {\bf Con 5} The connectivity of pipe components or members of steel frames assist the modeller in identifying all the connected components of a pipe spool or steel frame when using the outputs of this framework. Figure \ref{fig:PipeSpoolFrame} shows characteristic examples from the warehouse and the oil refinery datasets. The confidence level of the predicted class labels from the CLOI-NET-Class method is also an indicator of whether the performance of the instance segmentation under-segments instances. Figure \ref{fig:PipeSpoolFrame}(aiii) shows that the elbows of the pipe spool were predicted with uncertainty (confidence level score $\leq 80\%$) and this performance led to the under-segmentation of the pipe spool into cylinder and elbow instances. In this case, under-segmentation can be helpful for the modellers since segmentation of the pipe spool into its parts will be an easier task to achieve. 

{\bf Lim 1} The \textit{CLOI} dataset, although the largest available dataset of TLS industrial point clouds, is not enough to fully validate the proposed framework. More industrial facility point clouds with various configurations are needed to enhance the statistical validity of the framework with an increased confidence level and decrease the bias between facilities especially for the \textit{CLOI} classes that are underrepresented in the dataset. As demonstrated in \cite{agapaki2020cloi} more data is not always beneficial, so careful experimental set-up should be conducted to alleviate from negatively impacting the performance. {\bf Lim 2} Manual annotation of TLS industrial point clouds according to the data preparation explained in the experiments section is an onerous task. In these efforts, an automated segmentation interface should be adopted to enable for easy generation of labelled class and instance point clusters. {\bf Lim 3} Finally, the framework is not designed to segment objects of the same geometric group, for instance pipes, conduits and circular hollow sections or further object types within the same class i.e. globe valves, gate valves. This could be an interesting direction for future research.

\subsection{DATA AVAILABILITY}

Some or all data, models, or code used during the study were provided by a third party. Direct requests for these materials may be made to the provider as indicated in the Acknowledgements.

\subsection{ACKNOWLEDGEMENTS}

We thank our colleague Graham Miatt, who has provided insight, expertise and data that greatly assisted this research. We also express our gratitude to Bob Flint from BP International Centre for Business and Technology (ICBT), who provided data for evaluation. The research leading to these results has received funding from the Engineering and Physical Sciences Research Council (EPSRC) and the US National Academy of Engineering (NAE). AVEVA Group Plc. and BP International Centre for Business and Technology (ICBT) partially sponsor this research under grant agreements RG83104 and RG90532 respectively. We gratefully acknowledge the collaboration of all academic and industrial project partners. Any opinions, findings and conclusions or recommendations expressed in this material are those of the authors and do not necessarily reflect the views of the institutes mentioned above.

%\begin{figure}[!ht]
%\centering
%\includegraphics[width=\textwidth]{Figures/Definitionsins.png}
%\caption{Definitions of (a) class segmentation, (b) instance segmentation and (c) object detection on buildings \cite{Armeni20163DSpaces}.}
%\label{fig:definitions}
%\end{figure}

%\begin{landscape}
\begin{figure}[!ht]
\centering
\includegraphics[width=\textwidth]{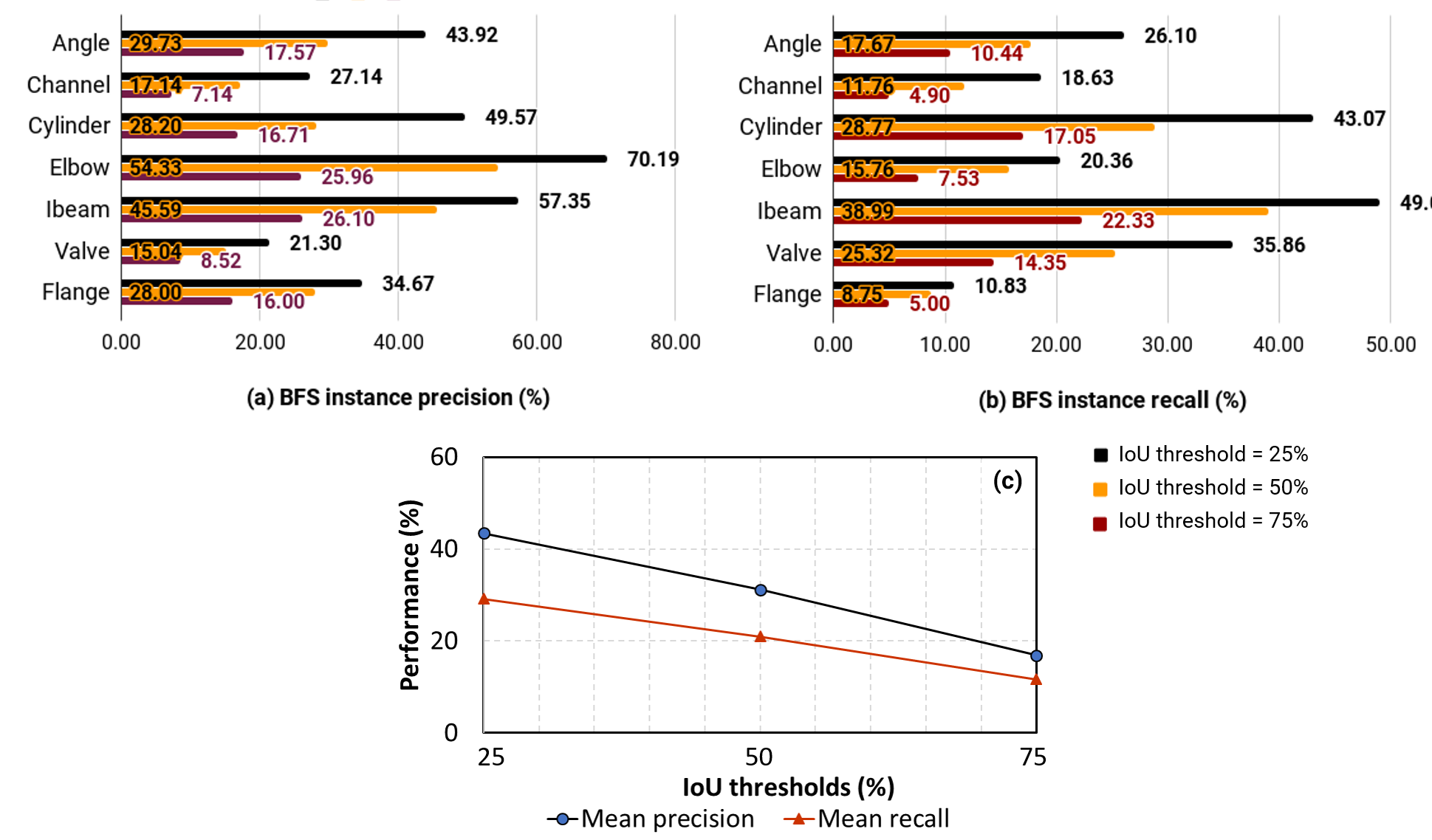}
\caption{(a) BFS instance precision and (b) recall per \textit{CLOI} class and (c) mean precision and recall for different IoU thresholds for the oil refinery facility.}
\label{fig:BFS_BP_pred}
\end{figure}

\begin{table}[!ht]
\centering
\caption{CLOI framework performance for the oil refinery dataset}
\large
\begin{tabular}{ l l l } 
\hline\hline
{\bf Method} & {\bf \makecell{mPrec \\ (\%)}} & {\bf \makecell{mRec \\ (\%)}} \\
\hline
%\ {\bf ASIS} \citep{Wang2019AssociativelyClouds} & 41.05 & 23.04 \\ 
%\ {\bf SGPN} \citep{Wang2018SGPN:Segmentation} & 35 & 19 \\ 
\ {\bf SGPN} \cite{Wang2018SGPN:Segmentation} & 5.3 & 6.5 \\
\ {\bf ASIS} \cite{Wang2019AssociativelyClouds} & 16.7 & 4.5 \\
\ {\bf CLOI-Framework (without boundary)} & 20.6 & 19.9 \\
\ {\bf CLOI-Framework} & 31.1 & 21.0 \\
\hline\hline
\end{tabular}
\label{table:inscomparisonexp}
\end{table}

\begin{table}[!ht]
\centering
\caption{Performance of instance segmentation networks per \textit{CLOI} shape in the oil refinery dataset}
\resizebox{\textwidth}{!}{%
    \begin{tabular}{ l l l l l l l l l }
    \hline\hline
    {\bf Prec (\%)} & {\bf Angles} & {\bf Channels} & {\bf Cylinders} & {\bf Elbows} & {\bf I-beams} & {\bf Valves} & {\bf Flanges} \\
    \hline
   % \ {\bf ASIS} & 57.4 & 17.4 & 51.2 & 58.3 & 66.1 & 30.1 & 23.9 & 24\\ 
    \ {\bf ASIS} & 0 & 0 & 27.2 & 25 & 41.5 & 6.3 & 0 \\ 
    \ {\bf SGPN} & 3.8 & 4.2 & 3.5 & 7.6 & 8.6 & 5.3 & 14 \\ 
    \ {\bf BFS} & 15.3 & 5.3 & 33.7 & 36.6 & 30 & 10.2 & 13.5 \\ 
    \ {\bf CLOI-Instance} & 29.7 & 17.1 & 28.2 & 54.3 & 45.6 & 15.1 & 28 \\
    \hline\hline
    {\bf Rec (\%)} & {\bf Angles} & {\bf Channels} & {\bf Cylinders} & {\bf Elbows} & {\bf I-beams} & {\bf Valves} & {\bf Flanges} \\
    \hline
    %\ {\bf ASIS} & 10.5 & 10.8 & 14.6 & 11 & 38.6 & 12.9 & 8.3 & 77.7\\ 
    \ {\bf ASIS} & 0 & 0 & 4.6 & 0.1 & 25.5 & 1.5 & 0 \\ 
    \ {\bf SGPN} & 2.8 & 3.5 & 4.2 & 3.6 & 5.9 & 15.2 & 4.6\\
    \ {\bf BFS} & 18.1 & 8.8 & 23.2 & 15 & 39.3 & 25.7 & 9.3 \\
    \ {\bf CLOI-Instance} & 17.7 & 11.7 & 28.8 & 15.7 & 39.0 & 25.3 & 8.8 \\ 
    \hline\hline
    \end{tabular}
}
\label{table:ASISSGPNexp}
\end{table}

\begin{table}[H]
\centering
\caption{Optimal class segmentation pre-annotation percentage of test facility data for active learning}
\large
\begin{tabular}{ l l } 
\hline\hline
{\bf Test facility} & {\bf \makecell{Optimal \\ pre-annotated data \\ (\%)}} \\
\hline
\ {\bf Warehouse} & 30 \\ 
\ {\bf Processing unit} & 30 \\
\ {\bf Oil refinery} & 25 \\
\ {\bf Petrochemical} & 20 \\
\hline\hline
\end{tabular}
\label{table:activeX}
\end{table}

\begin{table}[!ht]
\centering
\caption{Performance of the CLOI-Instance method per \textit{CLOI} shape for all the \textit{CLOI} datasets (IoU=25\%)}
\resizebox{\textwidth}{!}{%
    \begin{tabular}{ l l l l l l l l }
    \hline\hline
    {\bf Oil refinery} & {\bf Angles} & {\bf Channels} & {\bf Cylinders} & {\bf Elbows} & {\bf I-beams} & {\bf Valves} & {\bf Flanges} \\
    \hline
    \ {\bf Prec (\%)} & 43.9 & 27.1 & 49.6 & 70.2 & 57.4 & 21.3 & 34.7 \\
    \ {\bf Rec (\%)} & 26.1 & 18.6 & 43.1 & 20.4 & 49.1 & 35.9 & 10.8 \\ 
    \hline\hline
    {\bf Warehouse} & {\bf Angles} & {\bf Channels} & {\bf Cylinders} & {\bf Elbows} & {\bf I-beams} & {\bf Valves} & {\bf Flanges} \\
    \hline
    \ {\bf Prec (\%)} & 56 & 67.1 & 64.7 & 76.9 & 44.4 & 29.4 & 30.8 \\
    \ {\bf Rec (\%)} & 16.5 & 34.6 & 49.1 & 18.6 & 100 & 41.7 & 28.6 \\ 
    \hline\hline
    {\bf Petrochemical} & {\bf Angles} & {\bf Channels} & {\bf Cylinders} & {\bf Elbows} & {\bf I-beams} & {\bf Valves} & {\bf Flanges} \\
    \hline
    \ {\bf Prec (\%)} & 50 & 52.6 & 51.1 & 70 & 77.8 & 29.7 & 40 \\
    \ {\bf Rec (\%)} & 35 & 46.2 & 48.2 & 20 & 61.8 & 91.7 & 8.3 \\ 
    \hline\hline
    {\bf Processing unit} & {\bf Angles} & {\bf Channels} & {\bf Cylinders} & {\bf Elbows} & {\bf I-beams} & {\bf Valves} & {\bf Flanges} \\
    \hline
    \ {\bf Prec (\%)} & 36.8 & 39.1 & 48.8 & 50 & 72.3 & 41.4 & 14.3 \\
    \ {\bf Rec (\%)} & 8.7 & 23.7 & 35.5 & 9.1 & 46.4 & 43.5 & 0.5 \\
    \hline\hline
    \end{tabular}
}
\label{table:CLOIins}
\end{table}

%\begin{landscape}
\begin{table}[!ht]
\centering
\caption{Manual labour hours and total segmentation savings of the overall framework per \textit{CLOI} facility.}
%\resizebox{\textwidth}{!}{%
    \begin{tabular}{ l l l l l l l l l } 
    \hline\hline
    {\bf Oil refinery} & Angles & Channels & Cylinders & Elbows & I-beams & Valves & Flanges & Other\\
    \hline
    \ Recall (\%) & 26 & 19 & 43 & 20 & 49 & 36 & 11 & 25\\ 
    \ Total \# of shapes & 211 & 2347 & 94 & 121 & 723 & 215 & 202 & 563\\ 
    \ Manually segmented \\ \# of shapes & 156 & 1910 & 54 & 96 & 368 & 138 & 180 & 425\\
    \ Total \# of man hours & & & & 173 & & & & \\
    \ Total savings (\%) & & & & {\bf 26} & & & &\\
    \hline\hline
    {\bf Warehouse} & Angles & Channels & Cylinders & Elbows & I-beams & Valves & Flanges & Other\\
    \hline
    \ Recall (\%) & 16.5 & 34.6 & 56 & 18.6 & 100 & 41.7 & 28.6 & 27.9\\ 
    \ Total \# of shapes & 111 & 168 & 910 & 258 & 12 & 85 & 21 & 195\\ 
    \ Manually segmented \\ \# of shapes & 93 & 110 & 400 & 210 & 0 & 50 & 15 & 141\\
    \ Total \# of man hours & & & & 67 & & & &\\
    \ Total savings (\%) & & & & {\bf 42} & & & &\\
    \hline\hline
    {\bf Petrochemical} & Angles & Channels & Cylinders & Elbows & I-beams & Valves & Flanges & Other\\
    \hline
    \ Recall (\%) & 35 & 46.2 & 41.8 & 20 & 61.8 & 91.7 & 8.3 & 29\\ 
    \ Total \# of shapes & 60 & 264 & 1489 & 376 & 140 & 53 & 130 & 828\\
    \ Manually segmented \\ \# of shapes & 39 & 142 & 866 & 301 & 54 & 4 & 119 & 588\\
    \ Total \# of man hours & & & & 74 & & & & \\
    \ Total savings (\%) & & & & {\bf 37} & & & &\\
    \hline\hline
    {\bf Processing unit} & Angles & Channels & Cylinders & Elbows & I-beams & Valves & Flanges & Other\\
    \hline
    \ Recall (\%) & 8.7 & 23.7 & 35.5 & 9.1 & 46.4 & 43.5 & 0.4 & 25.1\\ 
    \ Total number of shapes & 188 & 34 & 1100 & 382 & 274 & 341 & 229 & 370\\ 
    \ Manually segmented \\ \# of shapes & 172 & 26 & 710 & 347 & 147 & 193 & 228 & 277\\
    \ Total \# of man hours & & & & 117 & & & &\\
    \ Total savings (\%) & & & & {\bf 28} & & & &\\
    \hline\hline
    \end{tabular}
%}
\label{table:savingsInstance}
\end{table}
%\end{landscape}

\begin{table}[!ht]
\centering
\caption{Correctly predicted cylinders of the petrochemical plant and warehouse point clouds using EdgeWise and our framework.}
%\resizebox{\textwidth}{!}{%
    \begin{tabular}{ l l l } 
    \hline\hline
    {\bf \makecell{\# of cylinders \\ correctly predicted}} & {\bf Warehouse} & {\bf Petrochemical} \\
    \hline
    \ EdgeWise & 468 & 164 \\ 
    \ Proposed framework & 510 & 623 \\ 
    \hline
    \hline
    \end{tabular}
%}
\label{table:cylindersComparison}
\end{table}

\begin{table}[!ht]
\centering
\caption{Percentage (\%) of the reduction of the labour hours of the \textit{CLOI} framework compared to EdgeWise per class.}
%\resizebox{\textwidth}{!}{%
    \begin{tabular}{ l l } 
    \hline\hline
    {\bf CLOI class} & {\bf \% of labour hour reduction} \\
    \hline
    \ Cylinders & 22.3 \\ 
    \ Channels & 40.4 \\
    \ I-beams & 81 \\
    \ Valves & 67 \\
    \ Elbows & 19.3 \\
    \ Flanges & 18.5 \\
    \ Angles & 25.7 \\
    \hline
    \hline
    \end{tabular}
%}
\label{table:savingsCLOI}
\end{table}

\nolinenumbers
\section{Appendix}
\label{appendix}

Appendix figures. 
\begin{figure}[!ht]
\centering
\includegraphics[width=\textwidth]{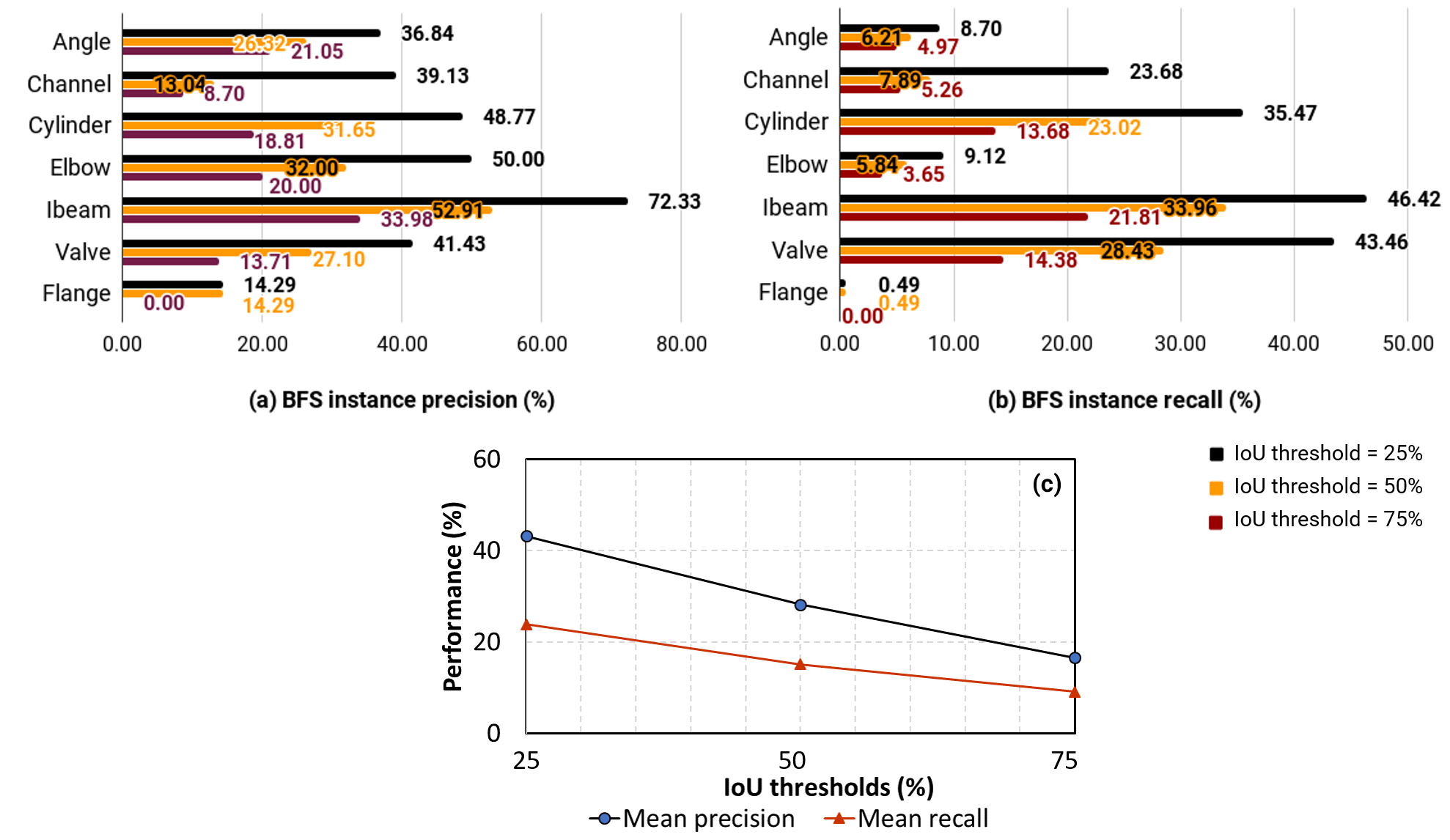}
\caption{(a) BFS instance precision and (b) recall per \textit{CLOI} class and (c) mean precision and recall for different IoU thresholds for the processing unit facility.}
\label{fig:BFS_NF_pred}
\end{figure}

\begin{figure}[!ht]
\centering
\includegraphics[width=\textwidth]{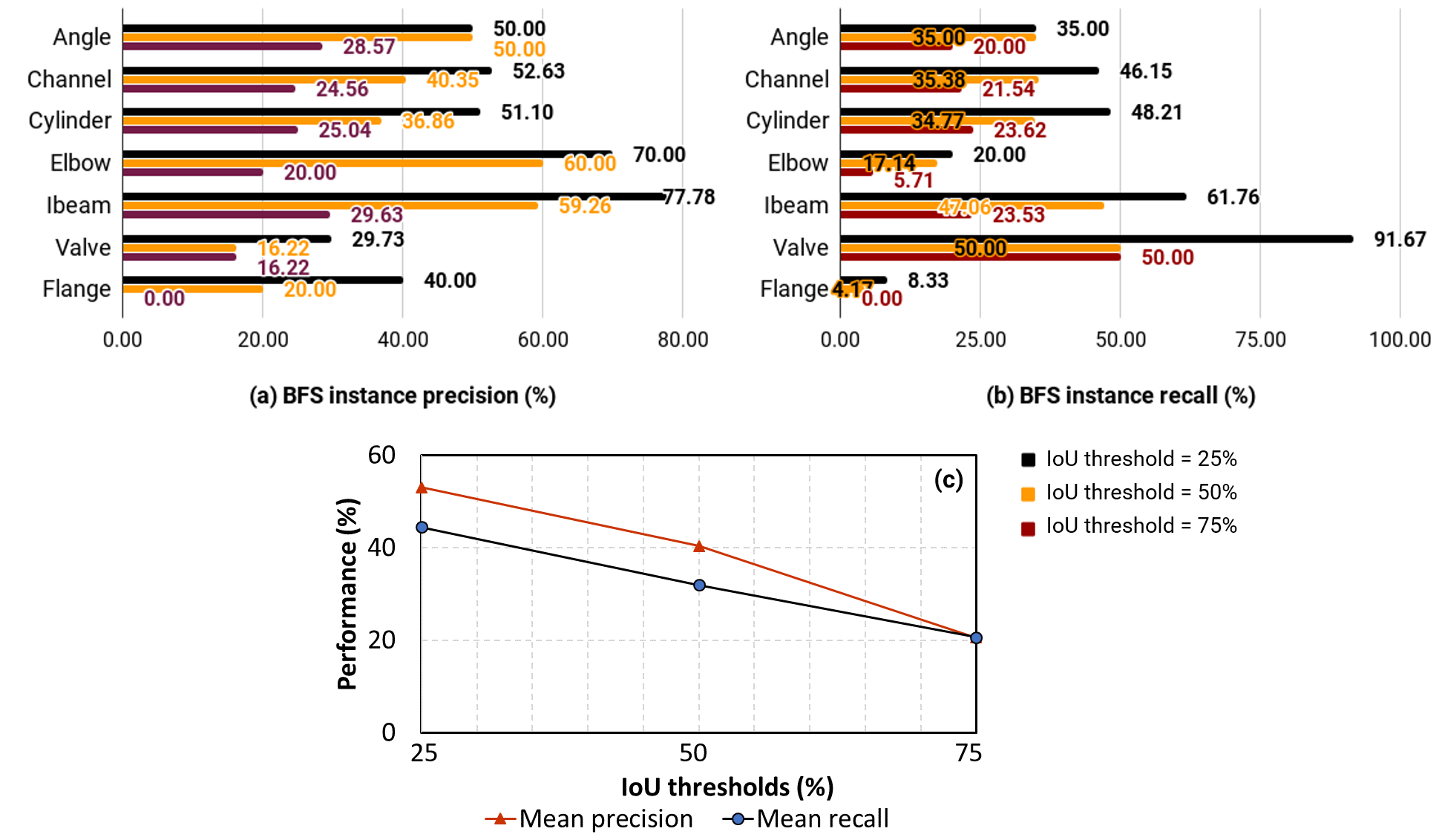}
\caption{(a) BFS instance precision and (b) recall per \textit{CLOI} class and (c) mean precision and recall for different IoU thresholds for the petrochemical plant facility.}
\label{fig:BFS_EATON_pred}
\end{figure}

\begin{figure}[!ht]
\centering
\includegraphics[width=\textwidth]{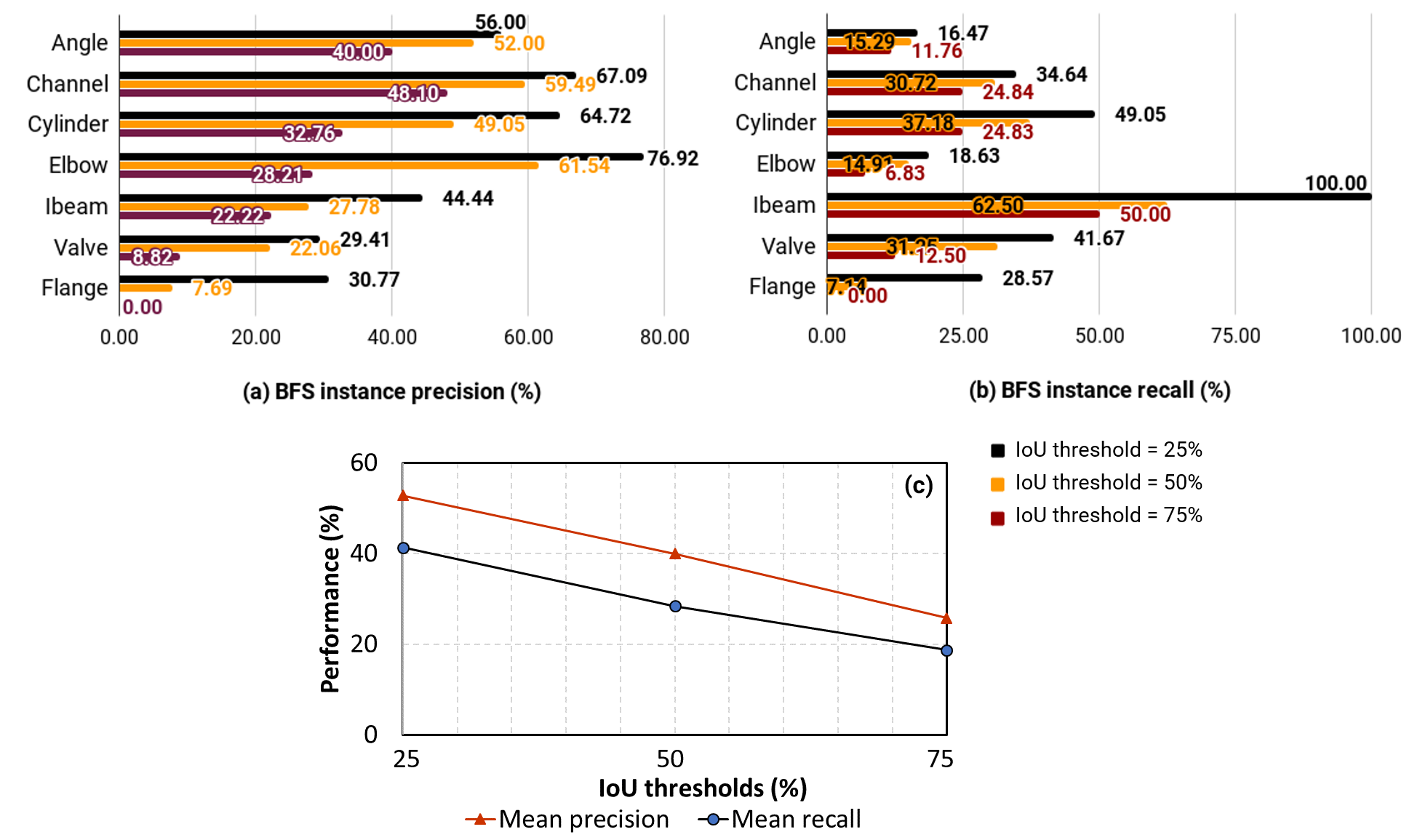}
\caption{(a) BFS instance precision and (b) recall per \textit{CLOI} class and (c) mean precision and recall for different IoU thresholds for the warehouse facility.}
\label{fig:BFS_THOR_pred}
\end{figure}

%\section{References, Citations and bibliographic entries}

\bibliography{references}

%
% Here's the list of references:
%
%\label{section:references}
%\bibliography{references}
%

\end{document}